\documentclass[12pt]{article}
\usepackage{latexsym,amssymb,amsfonts,amsmath}
\usepackage{mathrsfs}
\usepackage{graphicx}
\usepackage{color}
\usepackage{float}
\newlength{\extraspace}
\newlength{\extraspaces}
\newcommand{\be}{\begin{equation}
\addtolength{\abovedisplayskip}{\extraspaces}
\addtolength{\belowdisplayskip}{\extraspaces}
\addtolength{\abovedisplayshortskip}{\extraspace}
\addtolength{\belowdisplayshortskip}{\extraspace}}
\newcommand{\ee}{\end{equation}}

\newcommand{\bea}{\begin{eqnarray}
\addtolength{\abovedisplayskip}{\extraspaces}
\addtolength{\belowdisplayskip}{\extraspaces}
\addtolength{\abovedisplayshortskip}{\extraspace}
\addtolength{\belowdisplayshortskip}{\extraspace}}
\newcommand{\eea}{\end{eqnarray}}

\newcommand{\newsection}[1]{
\pagebreak[3]
\addtocounter{section}{1}
\setcounter{equation}{0}
\setcounter{subsection}{0}
\setcounter{footnote}{0}
\begin{flushleft}
{\large\bf \thesection. #1}
\end{flushleft}
\nopagebreak
\nopagebreak}

\newcommand{\newsubsection}[1]{
\vspace{5mm}
\pagebreak[3]
\addtocounter{subsection}{1}
\noindent{\bf \thesubsection. #1}
\nopagebreak
\vspace{2mm}
\nopagebreak}

\DeclareMathOperator{\Tr}{Tr} 
\DeclareMathOperator{\diag}{diag} 

\newcommand{\I}{\textbf{I}}

\newcommand{\bra}{\langle}
\newcommand{\ket}{\rangle}

\begin{document}

\addtolength{\baselineskip}{.8mm}

{\thispagestyle{empty}

\mbox{}                \hfill Revised: March 2020 \hspace{1cm}\\

\begin{center}
{\Large\bf Study of the interactions of the axion with mesons}\\
{\Large\bf and photons using a chiral effective Lagrangian model}\\
\vspace*{1.0cm}
{\large
Giacomo Landini\footnote{E-mail: giacomo.landini@phd.unipi.it}
and
Enrico Meggiolaro\footnote{E-mail: enrico.meggiolaro@unipi.it}
}\\
\vspace*{0.5cm}{\normalsize
{Dipartimento di Fisica, Universit\`a di Pisa,
and INFN, Sezione di Pisa,\\
Largo Pontecorvo 3, I-56127 Pisa, Italy}}\\
\vspace*{2cm}{\large \bf Abstract}
\end{center}

\noindent
We investigate the most interesting decay processes involving axions, photons and the lightest pseudoscalar mesons, making use of a chiral effective Lagrangian model with $L=3$ light quark flavors, which also includes the flavor-singlet pseudoscalar meson and implements the $U(1)$ axial anomaly of the fundamental theory. In particular, we compute the electromagnetic coupling of the axion to photons and we compare our result with the prediction of the Chiral Effective Lagrangian with $L=2$ light quark flavors. Moreover, we study the decay channels $\eta/\eta'\rightarrow \pi\pi a$ and we estimate the corresponding decay widths, using the existing bounds on the $U(1)_{PQ}$ breaking scale.
}

\newpage

\newsection{Introduction}

It is well known that the QCD Lagrangian $\mathcal{L}_{QCD}$ can be extended by adding the term $\mathcal{L}_{\theta}=\theta Q$, where $Q=\frac{g^{2}}{64\pi^{2}}\varepsilon^{\mu\nu\rho\sigma} G^{a}_{\mu\nu}G^{a}_{\rho\sigma}$ is the so-called \textit{topological charge density} and $\theta$ is a free parameter, which can assume any value in $[0,2\pi)$. This $\theta$-\textit{term} (or \textit{topological term}) introduces an explicit breaking of the CP symmetry in the strong sector (referred to as strong-CP violation). Despite the fact that $Q=\partial_{\mu}K^{\mu}$, where $K^{\mu}$ is the so-called \textit{Chern-Simons current}, its contribution is nonzero because of topologically nontrivial configurations of gauge fields, such as \textit{instantons}.
So far, however, no CP violation in the strong sector has been observed experimentally, constraining $\theta$ to be zero or extremely small. In particular, one can find a relationship between $\theta$ and the neutron electric-dipole moment \cite{Weinberg-book}, $d_{N}\simeq \frac{m_{\pi}^{2}}{m_{N}^{3}}e|\theta|\simeq 10^{-16} |\theta| \ e \cdot \text{cm}$, where $m_N$ is the neutron mass, whereas $m_{\pi}$ is the pion mass. From experimental data \cite{Neutron-EDM} we know that $d_{N}<10^{-26} \ e \cdot \text{cm}$, which leads to the upper bound $|\theta|<10^{-10}$.
(More refined relations among the neutron electric dipole moment and the $\theta$ angle and a more detailed discussion can be found in Refs. \cite{Baluni1979,CDVW1979,VP2009}; see also Ref. \cite{EDM-lattice} for a recent lattice determination).

This ``fine-tuning'' problem, known in the literature as the ``strong-CP problem'', is one of the open issues of the Standard Model. Among the several possible solutions, the most appealing is surely the one proposed by Peccei and Quinn (PQ) in 1977 \cite{PQ1977} and developed by Weinberg and Wilczek in 1978 \cite{Weinberg1978,Wilczek1978}.
The key idea (see also Ref. \cite{Peccei2008} for a recent review) is to extend the Standard Model by adding a new pseudoscalar particle, called ``axion'', in such a way that there is a new $U(1)$ global symmetry, referred to as $U(1)_{PQ}$, which is both spontaneously broken at a scale $f_a$ and anomalous (i.e., broken by quantum effects), with the related current satisfying the relation $\partial_{\mu}J^{\mu}_{PQ}=a_{PQ}Q$ , where $a_{PQ}$ is the so-called \textit{color anomaly} parameter.
The most general Lagrangian describing the QCD degrees of freedom and the axion has the following form:
\begin{equation} \label{axionlagrangian}
\mathcal{L} = \mathcal{L}_{QCD} + \frac{1}{2}\partial_{\mu}S_a\partial^{\mu}S_a - a_{PQ}\frac{S_a}{f_{a}}Q + \mathcal{L}_{int}[\partial_{\mu}S_a,\Psi] ,
\end{equation} 
where $S_a$ is the axion field, which under $U(1)_{PQ}$ transforms nonlinearly as
\begin{equation} \label{shiftaxion}
U(1)_{PQ}:~~~~ S_a\rightarrow S_a'= S_a + \gamma f_{a} .
\end{equation} 
The term $\mathcal{L}_{int}[\partial_{\mu}S_a,\Psi]$ describes the interactions between the axion and the quark fields and it is strongly model dependent.
The effect of this extension of the Lagrangian is to replace the static $\theta$ parameter of $\mathcal{L}_{QCD}$ with a dynamical degree of freedom, namely the combination $\theta-a_{PQ}\frac{S_a}{f_a}$: on the vacuum we get $\bra\theta-a_{PQ}\frac{S_a}{f_a}\ket=0$. Performing a $U(1)_{PQ}$ transformation \eqref{shiftaxion} with $\gamma=\frac{\theta}{a_{PQ}}$, we can rotate away the $\theta$ term, so obtaining a manifestly CP-conserving theory (with $\bra S_a \ket = 0$).

Moreover, it is well known that the $U(1)$ axial symmetry of QCD with $L$ light quark flavors (taken to be massless in the ideal \textit{chiral limit}; the physically relevant cases are $L=2$, with the quarks \textit{up} and \textit{down}, and $L=3$, including also the \textit{strange} quark),
\begin{equation} \label{U(1)_A}
U(1)_A:~~~~ q_i \rightarrow q'_i = e^{i\beta\gamma_5}q_i,~~~ i=1,\ldots,L ,
\end{equation}
is also anomalous, with the related $U(1)$ axial current $J^{\mu}_{5}=\bar{q}\gamma^{\mu}\gamma_{5}q$ satisfying the relation $\partial_{\mu}J^{\mu}_{5}=2LQ$.
Therefore, we find that the $U(1)_A \otimes U(1)_{PQ}$ transformations with the parameters $\beta$ and $\gamma$ satisfying the constraint $2L\beta + a_{PQ} \gamma = 0$, form a $U(1)$ subgroup which is spontaneously broken but anomaly-free (in the chiral limit): as a consequence, a new (pseudo-)Nambu-Goldstone boson appears in the spectrum, the axion.

In the original Peccei-Quinn-Weinberg-Wilczek (PQWW) model \cite{PQ1977,Weinberg1978,Wilczek1978} the scale $f_a$ was identified with the electroweak breaking scale $v \approx 250$ GeV, but this leads to large couplings between the axion and the Standard Model fields, which have been ruled out by experiments (see, for example, Ref. \cite{MS2015}).
In order to bypass these experimental bounds, the so-called ``invisible axion'' models were developed, such as the Kim-Shifman-Vainshtein-Zakharov (KSVZ) model \cite{Kim79,SVZ} and the Dine-Fischler-Srednicki-Zhitnisky (DFSZ) model \cite{DFS,Zhitnisky1980}, in which new heavy quarks or scalar fields, charged under $U(1)_{PQ}$ but neutral with respect to the Standard Model gauge group, are introduced. In these models, the $U(1)_{PQ}$ breaking scale $f_{a}$ is a free parameter of the theory and, assuming $f_{a} \gg v$, a very light axion with small couplings to the Standard Model fields is predicted, a scenario which is still compatible with the experimental bounds. At present, the more precise bounds on the $U(1)_{PQ}$ breaking scale come from astrophysical and cosmological considerations (see, for example, Ref. \cite{bounds}): $10^{9}~ \text{GeV} \lesssim f_a \lesssim 10^{17}~ \text{GeV}$.

All these models predict an axion-photon-photon coupling and therefore the electromagnetic decay of the axion in two photons: most of the experimental research concerning the axion is focused on this process (see, for example, Ref. \cite{IR2018} for an exhaustive review on both the theoretical aspects and the experimental research of axions and axion-like particles).
The electromagnetic interaction of the axion is usually parametrized as 
\begin{equation} \label{gammagammacoupling}
\Delta\mathcal{L}_{a\gamma\gamma} = -\frac{1}{4}g_{a\gamma\gamma}aF_{\mu\nu}\tilde{F}^{\mu\nu} ,
\end{equation}
where $a$ is the ``physical'' axion [as we will see, the field $S_a$, which appears in the Lagrangian \eqref{axionlagrangian}, has nonzero mixings with the QCD degrees of freedom, such as the pseudoscalar meson fields],
$F_{\mu\nu}$ is the electromagnetic field-strength tensor, $\tilde{F}^{\mu\nu}=\frac{1}{2}\varepsilon^{\mu\nu\rho\sigma}F_{\rho\sigma}$ is its dual, and $g_{a\gamma\gamma}$ is the axion-photon-photon coupling constant.
This last, in general, is the sum of two contributions, $g_{a\gamma\gamma} = g^0_{a\gamma\gamma} + g^{QCD}_{a\gamma\gamma}$, where $g^0_{a\gamma\gamma}$ is the model-dependent contribution proportional to the electromagnetic anomaly of the $U(1)_{PQ}$ symmetry (which can also be simply zero, as it happens in the original KSVZ model \cite{Kim79,SVZ}), while $g^{QCD}_{a\gamma\gamma}$ is the model-independent contribution coming from the minimal coupling to QCD (i.e., the mixing of the axion with the pseudoscalar mesons $\pi^0$, $\eta$, and $\eta'$).
The coupling constant $g^{QCD}_{a\gamma\gamma}$ has been computed using the Chiral Effective Lagrangian with $L=2$ light quark flavors both at the leading order (LO) in the momentum expansion [$\mathcal{O}(p^2)$] and at the next-to-leading order (NLO) [$\mathcal{O}(p^{4}$] (see Ref. \cite{GHVV2016} and references therein).

The aim of this paper is to compute the axion-photon-photon coupling constant $g^{QCD}_{a\gamma\gamma}$ and, moreover, to study the decay processes involving the axion and the lightest pseudoscalar mesons, making use of a chiral effective Lagrangian model proposed by Witten, Di Vecchia, Veneziano, \textit{et al.} \cite{Witten80,DV1980,etal80}, which describes the Nambu-Goldstone bosons originated by the spontaneous breaking of the $SU(3)_L\otimes SU(3)_R$ chiral symmetry (with $L=3$ light quark flavors) and the flavor-singlet pseudoscalar meson, implementing the $U(1)$ axial anomaly of the fundamental theory.

In Sec. 2, for the benefit of the reader, we briefly recall this chiral effective Lagrangian model, as well as its ``axionized'' version (see Ref. \cite{DS2014}).\\
Using this model, in Sec. 3 we compute the axion-photon-photon coupling constant $g^{QCD}_{a\gamma\gamma}$ and the result is compared with the one obtained using the Chiral Effective Lagrangian with $L=2$ light quark flavors.

Sec. 4 is devoted to the study of the hadronic decays $\eta/\eta'\rightarrow \pi\pi a$ (which, of course, cannot be studied using the Chiral Effective Lagrangian with $L=2$ light quark flavors, since the $\eta$ and $\eta'$ degrees of freedom are integrated out). Among all the possibile hadronic decays involving also the axion, these are the ones involving the lowest-energy hadrons.

In Sec. 5 we study the effects of a possible $U(1)$ axial condensate on the various quantities that we have evaluated in the previous sections: we do this by using a chiral effective Lagrangian model proposed in Ref. \cite{Meggiolaro1994} and then elaborated on in Refs. \cite{MM2003,Meggiolaro2011,MM2013}, which can be interpreted as an extension of the model considered in Sec. 2 with the inclusion of a $U(1)$ axial condensate.

In Sec. 6  we report numerical estimates for the axion-photon-photon coupling constant (making also a comparison with the prediction of the Chiral Effective Lagrangian with $L=2$ light quark flavors) and for the widths of the hadronic decays $\eta/\eta'\rightarrow \pi\pi a$.

Finally, in Sec. 7 we summarize and critically comment on the results that we have obtained in the previous sections for the electromagnetic and the hadronic processes involving the axion (considering the existing bounds on the $U(1)_{PQ}$ breaking scale) and we also give some prospects for further theoretical and experimental studies of the hadronic decays $\eta/\eta'\rightarrow \pi\pi a$.

\newsection{The effective Lagrangian model of Witten, Di Vecchia, Veneziano, \emph{et al.}, with the inclusion of the axion}

The effective Lagrangian model proposed by Witten, Di Vecchia, Veneziano, \textit{et al.} \cite{Witten80,DV1980,etal80} describes the Nambu-Goldstone bosons originated by the spontaneous breaking of the $SU(3)_L\otimes SU(3)_R$ chiral symmetry and the flavor-singlet pseudoscalar meson, implementing the $U(1)$ axial anomaly of the fundamental theory. We will refer to it as the ``WDV model''. Even if this model was derived and fully justified in the large-$N_c$ limit ($N_c$ being the number of colors), the numerical results obtained for the physical value $N_c=3$ turn out to be quite consistent with experimental data.
The Lagrangian is given by (see Ref. \cite{DV1980} for a detailed discussion)
\begin{equation}
\begin{split}
\mathcal{L}=&\frac{1}{2}\Tr\left[\partial_{\mu}U^{\dagger}\partial_{\mu}U\right]
+\frac{BF_{\pi}}{\sqrt{2}}\Tr\left[M(U+U^{\dagger})\right] \\ &+\frac{i}{2}Q\Tr[\ln U-\ln U^{\dagger}]+\frac{Q^{2}}{2A}+\theta Q .
\end{split}
\end{equation}
The mesonic field $U$ is represented by a $3\times 3$ complex matrix, which can be written in terms of the quark fields as $U_{ij} \sim \bar{q}_{jR}q_{iL}$, up to a multiplicative constant.\footnote{Throughout this paper, we shall use the following notations for the left-handed and right-handed quark fields: $q_{L,R} \equiv \frac{1}{2}(1 \pm \gamma_{5})q$, with $\gamma_5 \equiv -i\gamma^0\gamma^1\gamma^2\gamma^3$. Moreover, we shall adopt the convention $\varepsilon^{0123} = -\varepsilon_{0123} = 1$ for the (Minkowskian) completely antisymmetric tensor $\varepsilon^{\mu\nu\rho\sigma}$ ($=-\varepsilon_{\mu\nu\rho\sigma}$) which appears in the expressions of the topological charge density $Q$ and of the dual electromagnetic field-strength tensor $\tilde{F}^{\mu\nu}$.}
Under a general $SU(3)_L\otimes SU(3)_R\otimes U(1)_A$ transformation [$q_L \to q'_L = \widetilde{V}_L q_L$ and $q_R \to q'_R = \widetilde{V}_R q_R$, where $\widetilde{V}_{L}=e^{i\beta}V_L$, $\widetilde{V}_{R}=e^{-i\beta}V_R$, with ${V}_{L,R}\in SU(3)$] the field $U$ transforms as
\begin{equation} \label{transformation}
U \rightarrow U' = \widetilde{V}_{L}U\widetilde{V}_{R}^{\dagger} ,
\end{equation}
At zero temperature (after integrating out the scalar meson fields) we can adopt the usual nonlinear parametrization:
\begin{equation} \label{field}
U(x)=\frac{F_\pi}{\sqrt{2}}e^{\frac{i}{F_{\pi}}\left(\sum_{a=1}^{8}\pi_{a}(x)\lambda_{a}+\sqrt{\frac{2}{3}}S(x)\I\right)} ,
\end{equation}
where $\lambda_a ~(a=1,\ldots,8)$ are the usual generators of $SU(3)$ (\textit{Gell-Mann matrices}), normalized so as $\Tr\left[\lambda_a\lambda_b\right]=2\delta_{ab}$, and $\pi_{a}(x) $ are the nonsinglet pseudoscalar-meson fields,
while $S(x)$ is the flavor-singlet pseudoscalar-meson field. Moreover:
\begin{itemize}
\item $F_\pi$ is the pion decay constant.
\item $M=\text{diag}(m_u,m_d,m_s)$ is the quark-mass matrix.
\item B is a constant (with the dimension of a mass) which relates the squared masses of the pseudoscalar mesons and the quark masses: for example, $m_{\pi}^{2}=B(m_{u}+m_{d})$.
\end{itemize}
The topological charge density $Q$ is introduced as an auxiliary field, whereas $A$ is a parameter which (at least in the large-$N_c$ limit) can be identified with the topological susceptibility in the pure Yang-Mills theory ($A = -i \int d^4 x \langle T Q(x) Q(0) \rangle\vert_{YM}$). It is easy to see that the anomalous term $\Delta\mathcal{L}_{\text{anomaly}}=\frac{i}{2}Q\Tr[\ln U-\ln U^{\dagger}]$ is invariant under $SU(3)_L\otimes SU(3)_R$, while under $U(1)_A$ ($U \to U' = e^{2i\beta}U$) it transforms as
\begin{equation} \label{anomalylag}
\Delta\mathcal{L}_{\text{anomaly}}\rightarrow \Delta\mathcal{L}_{\text{anomaly}}-6\beta Q ,
\end{equation}
so correctly reproducing the $U(1)$ axial anomaly of the fundamental theory.

The model can be easily ``axionized'' (see Ref. \cite{DS2014}), essentially by promoting the parameter $\theta$ to the axion field $S_a$ (apart from a multiplicative constant $a_{PQ}/f_a$) and adding a kinetic term for it, i.e.,
\begin{equation}
\begin{split}
\mathcal{L}=&\frac{1}{2}\Tr\left[\partial_{\mu}U^{\dagger}\partial_{\mu}U\right]+\frac{1}{2}\partial_{\mu}N^{\dagger}\partial_{\mu}N 
+\frac{BF_{\pi}}{\sqrt{2}}\Tr\left[M(U+U^{\dagger})\right] \\ &+\frac{i}{2}Q\left\{\Tr[\ln U-\ln U^{\dagger}]+a_{PQ}[\ln N-\ln N^{\dagger}]\right\}+\frac{Q^{2}}{2A} ,
\end{split}
\end{equation}
where $N=f_a e^{i\frac{S_a}{f_a}}$ parametrizes the axion field in the standard notation for Nambu-Goldstone bosons.
It is convenient to integrate out the auxiliary field $Q$ using its equations of motion:
\begin{equation}
Q=-\frac{i}{2}A\left\{\Tr[\ln U-\ln U^{\dagger}]+a_{PQ}(\ln N-\ln N^{\dagger})\right\} .
\end{equation}
The resulting Lagrangian is given by
\begin{equation} \label{wdv}
\begin{split}
\mathcal{L}=&\frac{1}{2}\Tr[\partial_{\mu}U^{\dagger}\partial_{\mu}U]+\frac{1}{2}\partial_{\mu}N^{\dagger}\partial_{\mu}N+ \frac{BF_{\pi}}{\sqrt{2}}\Tr[M(U+U^{\dagger})] \\ &+\frac{A}{8}\left\{\Tr\left[\ln U-\ln U^{\dagger}\right]+a_{PQ}\left[\ln N-\ln N^{\dagger}\right]\right\}^2 .
\end{split}
\end{equation}
Expanding the Lagrangian up to the second order in the fields, we get the following squared-mass matrix for the fields $\pi_{3}$, $\pi_{8}$, $S$, $S_a$ (the mass term for the fields $\pi_1$, $\pi_2$, $\pi_4$, $\pi_5$, $\pi_6$, and $\pi_7$ being diagonal):
\begin{equation} \label{massmatrix}
\mathcal{M}^{2}=
\bigg .\left(
\begin{array}{cccc}
2B\tilde{m} & \frac{1}{\sqrt{3}}B\Delta & \sqrt{\frac{2}{3}}B\Delta & 0 \\
\frac{1}{\sqrt{3}}B\Delta & \frac{2}{3}B(\tilde{m}+2m_{s}) & \frac{2\sqrt{2}}{3}B(\tilde{m}-m_{s}) & 0 \\
\sqrt{\frac{2}{3}}B\Delta & \frac{2\sqrt{2}}{3}B(\tilde{m}-m_{s}) & \frac{2}{3}B(2\tilde{m}+m_{s})+\frac{6A}{F_{\pi}^{2}} & \frac{2\sqrt{3}Ab}{F_{\pi}^{2}} \\
0 & 0 & \frac{2\sqrt{3}Ab}{F_{\pi}^{2}} & \frac{2Ab^{2}}{F_{\pi}^{2}}
\end{array}
\bigg .\right) ,
\end{equation}
where we have defined
\begin{equation}
\tilde{m} \equiv \frac{1}{2}(m_{u}+m_{d}), \ \ \
\Delta \equiv m_{u}-m_{d}, \ \ \
b \equiv \frac{a_{PQ} F_\pi}{\sqrt{2} f_a} .
\end{equation}
The fields $\pi_{3}$, $\pi_{8}$, $S$, $S_a$ can be written in terms of the ``physical'' fields $\pi^0$, $\eta$, $\eta'$, $a$, associated with the mass eigenstates of Eq. \eqref{massmatrix}, as follows:
\begin{equation} \label{mixing}
\bigg .\left(
\begin{array}{c}
\pi_{3} \\
\pi_{8} \\
S \\
S_a \\
\end{array}
\bigg .\right)
=
\bigg .\left(
\begin{array}{cccc}
\theta_{\pi_{3}\pi_{3}} & \theta_{\pi_{3}\pi_{8}} & \theta_{\pi_{3}S} & \theta_{\pi_{3}S_a} \\
\theta_{\pi_{8}\pi_{3}} & \theta_{\pi_{8}\pi_{8}} & \theta_{\pi_{8}S} & \theta_{\pi_{8}S_a} \\
\theta_{S\pi_{3}} & \theta_{S\pi_{8}} & \theta_{SS} & \theta_{SS_a} \\
\theta_{S_a\pi_{3}} & \theta_{S_a\pi_{8}} & \theta_{S_aS} & \theta_{S_aS_a}
\end{array}
\bigg .\right)
\bigg .\left(
\begin{array}{c}
\pi^{0} \\
\eta \\
\eta' \\
a
\end{array}
\bigg .\right) ,
\end{equation}
where $\theta_{ij}$ is an orthogonal mixing matrix. 

From the astrophysical bounds on the scale $f_a$ \cite{bounds} (or better on $f_a/a_{PQ}$, but $a_{PQ} \sim \mathcal{O}(1)$ for the more realistic axion models \cite{DiLuzio2017}) we have: $10^{-18} \lesssim b \lesssim 10^{-10}$. As a consequence, it is surely legitimate to perform the computations only at the leading order in $b$.
In particular, diagonalizing the squared-mass matrix, we can derive the following expression (at the leading order in $b$) for the squared mass of the axion \cite{DS2014}:
\begin{equation}\label{axionmass} 
m_{a}^{2}=2b^{2}B\frac{m_{u}m_{d}m_{s}}{m_{u}m_{d}+m_{u}m_{s}+m_{d}m_{s}+\frac{BF_{\pi}^{2}}{A}m_{u}m_{d}m_{s}} .
\end{equation}
This expression is in perfect agreement with the well-known relationship (valid at the leading order in $b$) \cite{SVZ}: $m^{2}_a=\frac{2b^{2}}{F_\pi^{2}}\chi_{QCD}$, between the squared mass of the axion and the topological susceptibility of QCD, $\chi_{QCD}\equiv -i \int d^4 x \langle T Q(x) Q(0) \rangle\vert_{QCD}$, considering the expression of $\chi_{QCD}$ which is found using the WDV model (see Refs. \cite{DS2014,LM2018} and references therein).

\newsection{Electromagnetic decay of the axion}

In order to investigate the electromagnetic decay of the axion, we have to introduce the electromagnetic interactions into the Lagrangian \eqref{wdv}. This is done by (i) replacing the derivative of the field $U$ with the corresponding covariant derivative $D_\mu U = \partial_{\mu}U + i e A_\mu[\mathcal{Q},U]$, where $A_\mu$ is the electromagnetic field and $\mathcal{Q}=\diag(2/3,-1/3,-1/3)$ is the quark electric-charge matrix (in units of $e$, the absolute value of the electron charge), and (ii) by adding the following term, which reproduces the electromagnetic anomaly of the $U(1)$ and $SU(3)$ axial currents (see Ref. \cite{DNPV1981}):
\begin{equation}
\Delta\mathcal{L}_{\text{anomaly}}^{\text{(e.m.)}}=\frac{i}{2}G\Tr[\mathcal{Q}^{2}\left(\ln U -\ln U^{\dagger}\right)] ,
\end{equation}
where $G=\frac{e^{2}N_{C}}{32\pi^{2}}\varepsilon^{\mu\nu\rho\sigma}F_{\mu\nu}F_{\rho\sigma}$, $F_{\mu\nu}$ being the electromagnetic field-strenght tensor.
Using Eq. \eqref{field}, this term can be rewritten explicitly in terms of the meson fields, as follows:
\begin{equation}
\Delta\mathcal{L}^{\text{(e.m.)}}_{\text{anomaly}}=-\frac{G}{3F_{\pi}}\left(\pi_{3}+\frac{1}{\sqrt{3}}\pi_{8}+\frac{2\sqrt{2}}{\sqrt{3}}S\right) .
\end{equation}
Making use of Eq. \eqref{mixing}, one immediatly sees that this term contains an axion-photon-photon interaction of the type \eqref{gammagammacoupling}, with the following expression for the axion-photon-photon coupling constant:\footnote{As we have already said in the Introduction, this is indeed the model-independent contribution $g^{QCD}_{a\gamma\gamma}$ coming from the minimal coupling to QCD: for simplicity, in the rest of the paper we will refer to it simply as $g_{a\gamma\gamma}$, ignoring the model-dependent contribution $g^0_{a\gamma\gamma}$ proportional to the electromagnetic anomaly of the $U(1)_{PQ}$ symmetry.}
\begin{equation} \label{gammagamma}
g_{a\gamma\gamma}= \frac{\alpha_{\text{e.m.}}}{\pi F_{\pi}}\left(\theta_{\pi_{3}S_a}+\frac{1}{\sqrt{3}}\theta_{\pi_{8}S_a}+\frac{2\sqrt{2}}{\sqrt{3}}\theta_{SS_a}\right) ,
\end{equation}
where $\alpha_{\text{e.m.}} = \frac{e^2}{4\pi} \simeq \frac{1}{137}$ is the fine-structure constant.

To find the mixing parameters in Eq. \eqref{mixing}, we have to solve the equations for the eigenvectors of the matrix \eqref{massmatrix}. In particular, using the following notation:
\begin{equation}
|\pi_{3}\ket=
\bigg .\left(
\begin{array}{c}
1 \\
0 \\
0 \\
0
\end{array}
\bigg .\right) ,
\ \ \
|\pi_{8}\ket=
\bigg .\left(
\begin{array}{c}
0 \\
1 \\
0 \\
0
\end{array}
\bigg .\right) ,
\ \ \
|S\ket=
\bigg .\left(
\begin{array}{c}
0 \\
0 \\
1 \\
0
\end{array}
\bigg .\right) ,
\ \ \
|S_a\ket=
{
\bigg .\left(
\begin{array}{c}
0 \\
0 \\
0 \\
1
\end{array}
\bigg .\right)} ,
\end{equation}
the axion eigenvector is given by
\begin{equation} 
|{a}\ket=\theta_{\pi_{3}S_a}|{\pi_{3}}\ket+\theta_{\pi_{8}S_a}|{\pi_{8}}\ket+\theta_{SS_a}|{S}\ket+\theta_{S_aS_a}|{S_a}\ket =
{
\bigg .\left(
\begin{array}{c}
\theta_{\pi_3 S_a} \\
\theta_{\pi_8 S_a} \\
\theta_{S S_a} \\
\theta_{S_a S_a}
\end{array}
\bigg .\right)} .
\end{equation}
First we shall derive our expressions for $\Delta=0$, i.e., neglecting the experimentally small violations of the $SU(2)_V$ isospin symmetry. For $\Delta=0$ the mass matrix becomes diagonal with respect to $\pi_3$, which can thus be identified with $\pi^{0}$: therefore $\theta_{\pi_{3} S_a}|_{\Delta=0} = 0$. The eigenvector equations are, in this case:
\begin{equation} 
\begin{cases}
\left[\frac{2}{3}B(\tilde{m}+2m_{s})-m^{2}_{a}|_{\Delta=0}\right]\theta_{\pi_{8}S_a} +\frac{2\sqrt{2}}{3}B(\tilde{m}-m_{s})\theta_{SS_a} = 0 ,\\
\frac{2\sqrt{2}}{3}B(\tilde{m}-m_{s})\theta_{\pi_{8}S_a} + \left[\frac{2}{3}B(2\tilde{m}+m_{s})+\frac{6A}{F_{\pi}^{2}}-m^{2}_{a}|_{\Delta=0}\right]\theta_{SS_a}+2\sqrt{3}\frac{bA}{F_{\pi}^{2}}\theta_{S_aS_a} = 0 ,\\
\theta_{\pi_{3}S_a}^{2}+\theta_{\pi_{8}S_a}^{2}+\theta_{SS_a}^{2}+\theta_{S_aS_a}^{2} = 1 ,
\end{cases}
\end{equation}
where $m^{2}_{a}|_{\Delta=0}$ is given by the expression \eqref{axionmass} with $\Delta=0$, i.e., with $m_u=m_d=\tilde{m}$, and the third equation is the normalization condition. At the leading order in $b$, the following results are found:
\begin{eqnarray} \label{teta8}
\theta_{\pi_{8}S_a} = -\sqrt{\frac{2}{3}}b\left(\frac{m_{s}-\tilde{m}}{\tilde{m}+2m_{s}+\frac{BF_{\pi}^{2}}{A}\tilde{m}m_{s}}\right),~~
\theta_{SS_a}=-\frac{b}{\sqrt{3}}\left(\frac{\tilde{m}+2m_{s}}{\tilde{m}+2m_{s}+\frac{BF_{\pi}^{2}}{A}\tilde{m}m_{s}}\right), \nonumber \\
\theta_{S_aS_a} = 1.~~~~~~~~~~~~~~~~~~~~~~~~~~~~~~~~~~~~~~~~~~~~~~~~~~~~~~
\end{eqnarray}
Let's now consider the realistic case $\Delta\neq0$. If we write the squared-mass matrix \eqref{massmatrix} as $\mathcal{M}^2=\mathcal{M}^2_{\Delta=0}+\delta\mathcal{M}^2_{\Delta}$, where
\begin{equation} 
\delta \mathcal{M}^2_{\Delta}=
\bigg .\left(
\begin{array}{cccc}
0 & \frac{1}{\sqrt{3}}B\Delta & \sqrt{\frac{2}{3}}B\Delta & 0 \\
\frac{1}{\sqrt{3}}B\Delta & 0 & 0 & 0 \\
\sqrt{\frac{2}{3}}B\Delta & 0 & 0 & 0 \\
0 & 0 &0 & 0
\end{array}
\bigg .\right) ,
\end{equation} 
 we can evaluate the eigenvalues and the eigenstates of the matrix $\mathcal{M}^2$ at the first order in the parameter $\Delta$, by treating the term $\delta \mathcal{M}^2_{\Delta}$ as a small perturbation.
In particular, using first-order perturbation theory, we obtain for the axion eigenstate: $|a\ket=|a_{\Delta=0}\ket+|\delta a\ket$, with
\begin{equation} 
|{\delta a}\ket = \frac{1}{m^{2}_{a}|_{\Delta=0}-m^{2}_{\pi^{0}}|_{\Delta=0}}|{\pi^{0}_{\Delta=0}}\ket\bra{\pi^{0}_{\Delta=0}}|{\delta\mathcal{M}^2_\Delta}|{a_{\Delta=0}}\ket ,
\end{equation}
where $|\pi^{0}_{\Delta=0}\ket=|\pi_{3}\ket$ and $m_{\pi^{0}}^{2}|_{\Delta=0}=2B\tilde{m}$. Therefore, at the leading order in $\Delta$ and $b$: 
\begin{equation} \label{teta3}
\theta_{\pi_{3}S_a}=\frac{b\Delta}{\sqrt{2}\tilde{m}}\left(\frac{m_{s}}{\tilde{m}+2m_{s}+\frac{BF_{\pi}^{2}}{A}\tilde{m}m_{s}}\right) ,
\end{equation}
while the corrections to the other mixing parameters are of order $\mathcal{O}(\Delta^{2})$.\\
Finally, substituting the expressions \eqref{teta8} and \eqref{teta3} into Eq. \eqref{gammagamma}, we find the following result:
\begin{equation} \label{axioncoupling}
g_{a\gamma\gamma}=-\frac{\alpha_{\text{e.m.}}\sqrt{2}b}{3\pi F_{\pi}}\left(\frac{\tilde{m}+5m_{s}-\frac{3m_{s}\Delta}{2\tilde{m}}}{\tilde{m}+2m_{s}+\frac{BF_{\pi}^{2}}{A}\tilde{m}m_{s}}\right) .
\end{equation}
We observe that, if we take the formal limits $m_s\rightarrow\infty$ and $A\rightarrow\infty$, this result correctly reduces to the corresponding expression derived with the Chiral Effective Lagrangian ($\chi EL$) with $L=2$ flavors at LO, i.e., 
\begin{equation} \label{axioncouplingchi}
g_{a\gamma\gamma}|^{(\textrm{LO})}_{\chi EL}=-\frac{\alpha_{\text{e.m.}}\sqrt{2}b}{3\pi F_{\pi}}\left(\frac{m_{u}+4m_{d}}{m_{u}+m_{d}}\right) .
\end{equation}

\newsection{Hadronic decays with the axion}

This section is devoted to the study of the following processes:
\begin{equation} \label{decays}
\begin{cases}
\eta\rightarrow\pi^{0}+\pi^{0}+a ,\\
\eta\rightarrow\pi^{+}+\pi^{-}+a ,\\
\eta'\rightarrow\pi^{0}+\pi^{0}+a ,\\
\eta'\rightarrow\pi^{+}+\pi^{-}+a .\\
\end{cases}
\end{equation}
Among all the possible hadronic decays involving also the axion, these are the ones involving the lowest-energy hadrons. (Since, as we shall see below, every axion in the final or initial state implies a factor $b$ in the decay amplitude, multiaxion processes are extremely suppressed and we disregard them.)
The couplings of the axion with hadrons in general (and with the lightest mesons in particular) have already been investigated in the past literature, in many cases using also chiral effective Lagrangian techniques (see, e.g., Refs. \cite{couplings-with-Leff}), but never using the WDV Lagrangian \eqref{wdv}. Moreover, the particular processes \eqref{decays} have never been explicitly investigated before.\footnote{However, in the recent Ref. \cite{ASW2019} similar processes, such as $a\rightarrow 3\pi$ or $a\rightarrow \eta(\eta')\pi\pi$, involving QCD-scale \emph{axionlike} particles with masses $m_\pi \lesssim m_a \lesssim 3$ GeV, have been investigated, using also (for the case $m_a \lesssim 1$ GeV) chiral effective Lagrangian techniques.}
From an experimental point of view, there are well-known bounds on the decay widths of $\eta/\eta'\rightarrow \pi^{0}\pi^{0}$ and $\eta/\eta'\rightarrow \pi^{+}\pi^{-}$: they will be compared to our predictions in Sec. 7.

In order to compute the amplitudes of the processes \eqref{decays}, we must derive the interaction vertices between the axion and the pseudoscalar mesons. This can be achieved by expanding the WDV Lagrangian \eqref{wdv} up to the fourth order in the fields. We thus obtain the following quartic Lagrangian:
\begin{equation}
 \label{quartic}
\mathcal{L}_{4}=\frac{1}{4F_{\pi}^{2}}\left[-\frac{2}{3}f_{ijc}f_{c\alpha\beta}(\pi_{i}\partial_{\mu}\pi_{j})(\pi_{\alpha}\partial^{\mu}\pi_{\beta})\right]+\frac{B}{24F_{\pi}^{2}}\Tr\left[M\left(\displaystyle\sum_{a=1}^8 \pi_{a}\lambda_{a}+\sqrt{\frac{2}{3}}S~\I\right)^{4}\right] ,
\end{equation}
where $f_{abc}$ are the $SU(3)$ structure constants (defined as $[\lambda_{a},\lambda_b]=2if_{abc}\lambda_c$).
In particular, only the following term of the quartic Lagrangian is relevant for studying the processes \eqref{decays}:
\begin{equation} \label{decaysLag}
\Delta\mathcal{L}_{4}=\frac{B\tilde{m}}{3F_{\pi}^{2}}\left(\frac{1}{2}\pi_{3}^2+\pi^{+}\pi^{-}\right)\left(\pi_{8}^{2}+2S^{2}+2\sqrt{2}\pi_{8}S\right) ,
\end{equation}
where, as usual, $\pi^{\pm}=\frac{\pi_1\mp i\pi_2}{\sqrt{2}}$ are the charged pion fields.
As in the previous section, we shall work at the leading order in the parameter $b$. Moreover, considering also the explorative nature of this study, we shall neglect (for simplicity) isospin violations ($\Delta=0$).\footnote{This is usually expected to be a not too ``brutal'' approximation. For example, in the case of the axion-photon-photon coupling constant $g_{a\gamma\gamma}$ derived in the previous section, the percentage variation between the value obtained using Eq. \eqref{axioncoupling} and the corresponding value obtained by putting $\Delta = 0$ comes out to be about $17 \%$ (using the known values of the parameters that will be reported in Sec. 6).}
With these approximations, the following (relevant) mixing parameters are found diagonalizing the squared-mass matrix \eqref{massmatrix}:
\begin{equation} \label{mixing-parameters} 
\begin{cases}
\theta_{\pi_{3}\pi_{3}}=1,~~\theta_{\pi_{3}\pi_{8}}=\theta_{\pi_{3}S}=\theta_{\pi_{3}S_a}=0 ,\\
\theta_{\pi_8 \pi_3}=0,~~\theta_{\pi_{8}\pi_{8}}=\cos\varphi,~~
\theta_{\pi_{8}S}= -\sin\varphi,~~
\theta_{\pi_{8}S_a}=-\sqrt{\frac{2}{3}}b\left(\frac{m_{s}-\tilde{m}}{\tilde{m}+2m_{s}+\frac{B F_\pi^2}{A}\tilde{m}m_s}\right) ,\\
\theta_{S \pi_3}=0,~~\theta_{S\pi_{8}}=\sin\varphi,~~\theta_{S S}=\cos\varphi,~~\theta_{S S_a}=-\frac{b}{\sqrt{3}}\left(\frac{\tilde{m}+2m_s}{\tilde{m}+2m_{s}+\frac{B F_\pi^2}{A}\tilde{m}m_s}\right) ,
\end{cases}
\end{equation}
where $\varphi$ is the mixing angle between $\pi_8$ and $S$, given by \cite{Veneziano1979}:
\begin{equation} \label{mixing-angle}
\tan\varphi = \sqrt{2} - \frac{3}{2\sqrt{2}} \left[ \frac{m_\eta^2 - 2B\tilde{m}}{B(m_s-\tilde{m})} \right] .
\end{equation}
In particular, being $\Delta=0$, $\pi_{3}$ can be simply identified with $\pi^0$.
Making use of Eq. \eqref{mixing} and of the expressions \eqref{mixing-parameters}, the following quartic interaction terms are found from Eq. \eqref{decaysLag}:
\begin{equation}
\begin{cases}
\Delta\mathcal{L}_{\eta\pi^{0}\pi^{0}a}=\frac{1}{2}g_{\eta\pi^{0}\pi^{0}a}\eta(\pi^{0})^2 a ,\\
\Delta\mathcal{L}_{\eta\pi^{+}\pi^{-}a}=g_{\eta\pi^{+}\pi^{-}a}\eta \pi^{+}\pi^{-}a ,\\
\Delta\mathcal{L}_{\eta'\pi^{0}\pi^{0}a}=\frac{1}{2}g_{\eta'\pi^{0}\pi^{0}a}\eta'(\pi^{0})^2 a ,\\
\Delta\mathcal{L}_{\eta'\pi^{+}\pi^{-}a}=g_{\eta'\pi^{+}\pi^{-}a}\eta'\pi^{+}\pi^{-} a ,
\end{cases}
\end{equation}
with
\begin{equation} \label{couplingh1}
g_{\eta\pi^{0}\pi^{0}a}=g_{\eta\pi^{+}\pi^{-}a}\equiv-\frac{2\sqrt{2}bB}{\sqrt{3}F_{\pi}^{2}}\left(\cos\varphi+\sqrt{2}\sin\varphi\right)\left(\frac{\tilde{m}m_s}{\tilde{m}+2m_{s}+\frac{B F_\pi^2}{A}\tilde{m}m_s}\right) ,
\end{equation}
and
\begin{equation} \label{couplingh2}
g_{\eta'\pi^{0}\pi^{0}a}=g_{\eta'\pi^{+}\pi^{-}a}\equiv-\frac{2\sqrt{2}bB}{\sqrt{3}F_{\pi}^{2}}\left(-\sin\varphi+\sqrt{2}\cos\varphi\right)\left(\frac{\tilde{m}m_s}{\tilde{m}+2m_{s}+\frac{B F_\pi^2}{A}\tilde{m}m_s}\right) .
\end{equation}
The equality of the decay amplitudes $\mathcal{A}(\eta\rightarrow \pi^0\pi^0a)=g_{\eta\pi^{0}\pi^{0}a}$ and $\mathcal{A}(\eta\rightarrow \pi^+\pi^-a)=g_{\eta\pi^{+}\pi^{-}a}$ (as well as of the amplitudes $\mathcal{A}(\eta'\rightarrow \pi^0\pi^0a)=g_{\eta'\pi^{0}\pi^{0}a}$ and $\mathcal{A}(\eta'\rightarrow \pi^+\pi^-a)=g_{\eta'\pi^{+}\pi^{-}a}$) is, of course, a consequence of the fact that we are neglecting isospin violations.

\newsection{Effects of an extra $U(1)$ axial condensate}

In this section, we will try to answer the following question: considering the relevance of the $U(1)$ axial symmetry in defining the physical aspects of a hypothetical axion (i.e., its mass and its interactions), could a (no matter how small) hypothetical $U(1)$ axial condensate significantly modify these expectations?
More precisely, we will study the effects of a possible $U(1)$ axial condensate on the various quantities that we have evaluated in the previous sections, by using a chiral effective Lagrangian model proposed in Ref. \cite{Meggiolaro1994} and then elaborated on in Refs. \cite{MM2003,Meggiolaro2011,MM2013}: it can be interpreted as an extension of the WDV model with the inclusion of a $U(1)$ axial condensate and therefore we will refer to it as the ``extended model''.
In this model the $U(1)$ axial anomaly is implemented as in the WDV model (by properly introducing the auxiliary field $Q$), so that it correctly satisfies the transformation property \eqref{anomalylag} under the chiral group, but it also includes an extra $U(1)$ axial condensate, in addition to the usual chiral condensate $\bra\bar{q}q\ket$. This $U(1)$ axial condensate has the form $C_{U(1)} = \bra O_{U(1)}\ket$, where, for a theory with $L$ light quark flavors, $O_{U(1)}$ is a $2L$-quark local operator that has the chiral transformation properties of $O_{U(1)}\sim \det_{st} (q_{sR}q_{tL})+\det_{st} (q_{sL}q_{tR})$, where $s,t = 1,\ldots,L$ are flavor indices. The color indices (not explicitly indicated) are arranged in such a way that (i) $O_{U(1)}$ is a color singlet and (ii) $C_{U(1)}$ is a genuine 2$L$-quark condensate, i.e., it has no disconnected part proportional to some power of the quark-antiquark chiral condensate $\bra\bar{q}q\ket$. The explicit form of this condensate has been discussed in detail in Ref. \cite{Meggiolaro2011}. In what follows we shall consider the case $L=3$.

The Lagrangian of the extended model is thus written in terms of the topological charge density $Q$, the usual mesonic field $U_{ij}\sim \bar{q}_{jR}q_{iL}$, and a new field variable $X\sim \det_{st}\bar{q}_{sR}q_{tL}$, associated with the $U(1)$ axial condensate, which under a general $SU(3)_L \otimes SU(3)_R \otimes U(1)_A$ chiral transformation [see Eq. \eqref{transformation}] transforms as:
\begin{equation}
X\rightarrow \det(\tilde{V}_L)\det(\tilde{V}_R)^* X .
\end{equation}
In the usual nonlinear parametrization, the field X can be written as
\begin{equation}
X=\frac{F_X}{\sqrt{2}}e^{i\frac{\sqrt{2}}{F_X}S_X} ,
\end{equation}
where $F_X$ is essentially the vacuum expectation value of $X$ ($\bra X\ket=\frac{F_X}{\sqrt{2}}$), i.e., the $U(1)$ axial condensate, and $S_X$ is an exotic flavor-singlet pseudoscalar field. The model can be ``axionized'' in the same way as the WDV model.
The Lagrangian of the ``axionized extended model'' is written as:
\begin{equation}
\begin{split}
\mathcal{L}&=\frac{1}{2}\Tr[\partial_{\mu}U^{\dagger}\partial^{\mu}U]+ \frac{1}{2}\partial_{\mu}X^{\dagger}\partial^{\mu}X+\frac{1}{2}\partial_{\mu}N^{\dagger}\partial^{\mu}N \\
&+\frac{BF_{\pi}}{\sqrt{2}}\Tr\left[M(U+U^{\dagger})\right]+\frac{\kappa_1}{2\sqrt{2}}(X^{\dagger}\det U + X\det U^{\dagger})+\frac{Q^{2}}{2A} \\ 
&+\frac{i}{2}Q\left\{\omega_1\Tr[\ln U-\ln U^{\dagger}]+(1-\omega_1)(\ln X-\ln X^{\dagger})+a_{PQ}(\ln N-\ln N^{\dagger})\right\} .
\end{split}
\end{equation}
Integrating out the auxiliary field $Q$, one obtains:
\begin{equation}
\begin{split}
\mathcal{L}&=\frac{1}{2}\Tr[\partial_{\mu}U^{\dagger}\partial^{\mu}U]+ \frac{1}{2}\partial_{\mu}X^{\dagger}\partial^{\mu}X+\frac{1}{2}\partial_{\mu}N^{\dagger}\partial^{\mu}N \\
&+\frac{BF_{\pi}}{\sqrt{2}}\Tr\left[M(U+U^{\dagger})\right]+\frac{\kappa_1}{2\sqrt{2}}(X^{\dagger}\det U + X\det U^{\dagger}) \\ 
&+\frac{A}{8}\left\{\omega_1\Tr[\ln U-\ln U^{\dagger}]+(1-\omega_1)[\ln X-\ln X^{\dagger}]+a_{PQ}[\ln N-\ln N^{\dagger}]\right\}^{2} .
\end{split}
\end{equation}
The model is characterized, with respect to the WDV model, by three new parameters: $\omega_1$, $\kappa_1$, and $F_X$. As already observed in Refs. \cite{Meggiolaro2011,LM2018}, the Lagrangian of the extended model reduces to that of the WDV model by \emph{first} choosing $\omega_{1}=1$ and \emph{then} letting $F_X\rightarrow0$. Therefore, $\omega_{1}=1$ seems to be the most ``natural'' choice, at least at low temperatures, near $T=0$, where minimal deviations from the results of the WDV model are expected (on the other side, $\omega_{1}$ must necessarily vanish above the chiral transition temperature in order to avoid a singular behaviour of the anomalous term: see Refs. \cite{Meggiolaro1994,MM2013}).

Expanding the Lagrangian up to the second order in the fields, one finds the following squared-mass matrix for the fields $\pi_{3}$, $\pi_{8}$, $S$, $S_X$, $S_a$:
\begin{equation*} 
\mathcal{M}^{2}=
\bigg .\left(
\begin{array}{ccccc}
2B\tilde{m} & \frac{1}{\sqrt{3}}B\Delta & \sqrt{\frac{2}{3}}B\Delta & 0 & 0 \\
\frac{1}{\sqrt{3}}B\Delta & \frac{2}{3}B(\tilde{m}+2m_{s}) & \frac{2\sqrt{2}}{3}B(\tilde{m}-m_{s}) & 0 & 0 \\
\sqrt{\frac{2}{3}}B\Delta & \frac{2\sqrt{2}}{3}B(\tilde{m}-m_{s}) & \frac{2}{3}B(2\tilde{m}+m_{s})+\frac{6(A\omega_{1}^{2}+c)}{F_{\pi}^{2}} & \frac{2\sqrt{3}[A\omega_{1}(1-\omega_{1})-c]}{F_{\pi}F_{X}} & \frac{2\sqrt{3}bA\omega_{1}}{F_{\pi}^{2}} \\
0 & 0 & \frac{2\sqrt{3}[A\omega_{1}(1-\omega_{1})-c]}{F_{\pi}F_{X}} & \frac{2[A(1-\omega_{1})^{2}+c]}{F_{X}^{2}} & \frac{2bA(1-\omega_{1})}{F_{\pi}F_{X}} \\
0 & 0 & \frac{2\sqrt{3}bA\omega_{1}}{F_{\pi}^{2}} & \frac{2bA(1-\omega_{1})}{F_{\pi}F_{X}} & \frac{2b^{2}A}{F_{\pi}^{2}}
\end{array}
\bigg .\right) ,
\end{equation*}
where
\begin{equation*}
c \equiv \kappa_1\frac{F_X}{2}\left(\frac{F_\pi}{\sqrt{2}}\right)^{3} .
\end{equation*}
The eigenstates of this matrix are the usual pseudoscalar mesons $\pi^{0}$, $\eta$ and $\eta'$, plus another exotic pseudoscalar state, called $\eta_{X}$, and the axion. At the leading order in $b$, the following value for the squared mass of the axion is found:
\begin{equation} \label{EMmass}
m_{a}^{2}=2b^{2}B\frac{m_{u}m_{d}m_{s}}{m_{u}m_{d}+m_{u}m_{s}+m_{d}m_{s}+\frac{BF_{\pi}^{2}}{A}\left(1+\frac{A(1-\omega_{1})^{2}}{c}\right)m_{u}m_{d}m_{s}} .
\end{equation}
Also in this case [see the discussion after Eq. \eqref{axionmass}], the expression for $m_a^{2}$ turns out to be in agreement with the relation $m_a^{2}=\frac{2b^{2}}{F_\pi^{2}}\chi_{QCD}$, considering the expression of $\chi_{QCD}$ which is found using the extended model (see Ref. \cite{LM2018}). Moreover, we notice that for $\omega_{1}\neq1$ the mass of the axion in the extended model is smaller than the one obtained in the WDV model, due to the positive corrective factor in the denominator. If, instead, we consider $\omega_{1}=1$, the result coincides precisely with the result \eqref{axionmass} of the WDV model, independently of the other parameters ($\kappa_1$ and $F_X$) of the extended model. This is not a totally unexpected result since in this particular case the potential coincides with that of the WDV model, apart from a term independent of the axion field (a more detailed explanation of this can be found in Ref. \cite{LM2018}).

Concerning the axion-photon-photon coupling $g_{a\gamma\gamma}$, the following result is found:
\begin{equation} \label{EMcoupling}
g_{a\gamma\gamma}=-\frac{\alpha_{\text{e.m.}}\sqrt{2}b}{3\pi F_\pi}\left(\frac{\tilde{m}+5m_{s}-\frac{3m_{s}\Delta}{2\tilde{m}}}{\tilde{m}+2m_{s}}\right)\left\{\frac{A[\omega_{1}-(1-\omega_{1})z]}{BF_{\pi}^{2}\frac{\tilde{m}m_{s}}{\tilde{m}+2m_{s}}+A\omega_{1}^{2}+c+z\left[c-A\omega_{1}(1-\omega_{1})\right]}\right\} ,
\end{equation}
with $z \equiv \frac{A\omega_{1}(1-\omega_{1})-c}{{A(1-\omega_{1})^{2}}+c}$.
Also in this case, setting the ``natural'' value $\omega_{1}=1$ we recover the WDV expression \eqref{axioncoupling}, independently of the other parameters of the extended model.

Finally, we analyze the hadronic decays \eqref{decays} with the axion. The explicit computation shows that the quartic Lagrangian is the same of the WDV model [see Eq. \eqref{quartic}], apart from an additional term $\delta \mathcal{L}_4^{(c)}= \frac{c}{6}\left(\frac{\sqrt{3}S}{F_\pi}-\frac{S_X}{F_X}\right)^{4}$. Anyway, if we neglect the isospin violations ($\Delta=0$), this term does not contribute to the processes \eqref{decays}.
If, for the reasons explained above, we take $\omega_{1}=1$, which is the ``natural'' choice (at least at $T=0$), we easily derive (proceeding as in Sec. 4 and making use of the results already found in Refs. \cite{MM2003,Meggiolaro2011}) the following expressions for the coupling constants $g_{\eta\pi\pi a}$ and $g_{\eta'\pi\pi a}$:
\begin{equation} \label{EMcouplingh1}
g_{\eta\pi^{0}\pi^{0}a}=g_{\eta\pi^{+}\pi^{-}a}\equiv-\frac{2\sqrt{2}bB}{\sqrt{3}F_{\pi}^{2}}\left(\cos\tilde{\varphi}+\frac{\sqrt{2}F_\pi}{F_{\eta'}}\sin\tilde{\varphi}\right)\left(\frac{\tilde{m}m_s}{\tilde{m}+2m_{s}+\frac{B F_\pi^2}{A}\tilde{m}m_s}\right) ,
\end{equation}
and
\begin{equation} \label{EMcouplingh2}
g_{\eta'\pi^{0}\pi^{0}a}=g_{\eta'\pi^{+}\pi^{-}a}\equiv-\frac{2\sqrt{2}bB}{\sqrt{3}F_{\pi}^{2}}\left(-\sin\tilde{\varphi}+\frac{\sqrt{2}F_\pi}{F_{\eta'}}\cos\tilde{\varphi}\right)\left(\frac{\tilde{m}m_s}{\tilde{m}+2m_{s}+\frac{B F_\pi^2}{A}\tilde{m}m_s}\right) ,
\end{equation}
where $F_{\eta'} \simeq \sqrt{F_{\pi}^{2}+3F_{X}^{2}}$ can be interpreted as the $\eta'$ decay constant (see Refs. \cite{Meggiolaro1994,MM2003,Meggiolaro2011}) and $\tilde{\varphi}$ is the mixing angle between $\pi_8$ and $S$, which turns out to be a bit larger than the value $\varphi$ in Eq. \eqref{mixing-angle}, being \cite{MM2003,Meggiolaro2011}: $\tan\tilde{\varphi} = \frac{F_{\eta'}}{F_\pi} \tan\varphi$.
We observe that in the limit $F_X \rightarrow 0$ we have $F_{\eta'} \rightarrow F_\pi$ and $\tilde{\varphi} \rightarrow \varphi$, and the expressions \eqref{EMcouplingh1} and \eqref{EMcouplingh2} correctly reduce to the WDV results \eqref{couplingh1} and \eqref{couplingh2}.

Recalling the upper limit $|F_X| \lesssim 20$ MeV found in Refs. \cite{Meggiolaro1994,MM2003,Meggiolaro2011}, we have that $1 \leq F_{\eta'}/F_\pi \lesssim 1.07$. Using also the fact that the mixing angle is quite small [Eq. \eqref{mixing-angle} predicts a value $\varphi \simeq 5.5^\circ$ and thus $5.5^\circ \lesssim \tilde{\varphi} \lesssim 5.85^\circ$], we find that the coupling constant $g_{\eta\pi\pi a}$ [Eq. \eqref{EMcouplingh1}] practically coincides with the WDV result \eqref{couplingh1}, while the coupling constant $g_{\eta'\pi\pi a}$ [Eq. \eqref{EMcouplingh2}] approximately gets, with respect to the WDV result \eqref{couplingh2}, a multiplicative factor $0.94 \lesssim F_\pi/F_{\eta'} \leq 1$.

\newpage

\newsection{Numerical results}

In this section we report numerical estimates for the axion-photon-photon coupling constant and for the decay widths of the hadronic processes \eqref{decays} with the axion.
For the numerical computations, we have used the following values of the known parameters:
\begin{itemize}
\item $F_\pi=(92.1 \pm 1.2)$ MeV (see Ref. \cite{PDG}, where the value of $f_{\pi}\equiv\sqrt{2}F_{\pi}$ is reported).
\item $A=(180 \pm 5 ~\text{MeV})^4$ (see Ref. \cite{VP2009} and references therein).
\item For what concerns the parameter $B$ and the quark masses $m_{u}$, $m_{d}$, $m_{s}$, we can make use of the well-known relations (see, e.g., Ref. \cite{Weinberg-book}) between $Bm_u$, $Bm_d$, $Bm_s$ and the squared pseudoscalar-meson masses, derived using leading-order chiral perturbation theory (and ignoring small corrections due to the mixing with the axion):
\begin{equation} \label{masse}
\begin{cases}
Bm_{u} = m_{\pi^{0}}^{2}-\frac{1}{2}(m_{K^{0}}^{2}-m_{K^{+}}^{2}+m_{\pi^{+}}^{2}) ,\\
Bm_{d} = \frac{1}{2}(m_{K^{0}}^{2}-m_{K^{+}}^{2}+m_{\pi^{+}}^{2}) ,\\
Bm_{s} = \frac{1}{2}(m_{K^{0}}^{2}+m_{K^{+}}^{2}-m_{\pi^{+}}^{2}) .
\end{cases}
\end{equation}
The pseudoscalar-mesons masses are given by \cite{PDG}
\begin{equation} \label{masses}
\begin{cases}
m_{\pi^{+}}=139.57061(24) ~\text{MeV} ,\\
m_{\pi^{0}}=134.9770(5) ~\text{MeV} ,\\
m_{K^{+}}=493.677(16) ~\text{MeV} ,\\
m_{K^{0}}=497.611(13) ~\text{MeV} .
\end{cases}
\end{equation}
We also need $m_{\eta} = 547.862(17)$ MeV and $m_{\eta'} = 957.78(6)$ MeV.
\end{itemize}

\newsubsection{Axion-photon-photon coupling constant $g_{a\gamma\gamma}$}

In Table 1 we report the numerical estimate for the axion-photon-photon coupling constant $g_{a\gamma\gamma}$, obtained using the expression \eqref{axioncoupling} that we have derived in Sec. 3 using the ``axionized'' WDV model (as we have seen in Sec. 5, this expression is not modified using, in place of the WDV model, a ``natural'' extension of it which also includes an extra U(1) axial condensate): for comparison, we also report the corresponding estimates derived using the Chiral Effective Lagrangian ($\chi EL$) with $L=2$ light quark flavors at LO [$\mathcal{O}(p^{2})$] and NLO [$\mathcal{O}(p^{4})$] (see Ref. \cite{GHVV2016} and references therein).
\begin{table}[H]
\begin{center}
\begin{tabular}{| l | l |}
\hline
& $|g_{a\gamma\gamma}|/b$ [$10^{-5}\text{MeV}^{-1}$] \\ \hline
$\chi EL$ ($L=2$) at LO \cite{GHVV2016} & $3.59\pm0.05$ \\ \hline
$\chi EL$ ($L=2$) at NLO \cite{GHVV2016} & $3.42\pm0.07$ \\ \hline
WDV ($L=3$) [Eq. \eqref{axioncoupling}] & $3.29\pm0.06$ \\ \hline
\end{tabular}
\end{center}
\caption{Numerical values of the axion-photon-photon coupling constant $g_{a\gamma\gamma}$, obtained using Eq. \eqref{axioncoupling}, compared with the predictions of the Chiral Effective Lagrangian ($L=2$) at LO and NLO.}
\end{table}

\newsubsection{Hadronic decay widths with the axion}

The decay widths for the processes \eqref{decays} are given by
\begin{equation}
\begin{cases}
\Gamma(\eta\rightarrow\pi^{0}\pi^{0}a)=\frac{1}{2m_{\eta} \cdot 2!}|g_{\eta\pi^{0}\pi^{0}a}|^{2}\Phi^{(3)}(m_{\eta}|m_{\pi^{0}},m_{\pi^{0}},m_a) ,\\
\Gamma(\eta\rightarrow\pi^{+}\pi^{-}a)=\frac{1}{2m_{\eta}}|g_{\eta\pi^{+}\pi^{-}a}|^{2}\Phi^{(3)}(m_{\eta}|m_{\pi^{+}},m_{\pi^{-}},m_a) ,\\
\Gamma(\eta'\rightarrow\pi^{0}\pi^{0}a)=\frac{1}{2m_{\eta'} \cdot 2!}|g_{\eta'\pi^{0}\pi^{0}a}|^{2}\Phi^{(3)}(m_{\eta'}|m_{\pi^{0}},m_{\pi^{0}},m_a) ,\\
\Gamma(\eta'\rightarrow\pi^{+}\pi^{-}a)=\frac{1}{2m_{\eta'}}|g_{\eta'\pi^{+}\pi^{-}a}|^{2}\Phi^{(3)}(m_{\eta'}|m_{\pi^{+}},m_{\pi^{-}},m_a) ,
\end{cases}
\end{equation}
where the amplitudes $g_{\eta\pi\pi a}$ and $g_{\eta'\pi\pi a}$ are given by Eqs. \eqref{couplingh1} and \eqref{couplingh2} respectively and $\Phi^{(3)}(M|m_1,m_2,m_3)$ is the phase space (with the usual ``relativistic'' normalization) for three particles of masses $m_1$, $m_2$, $m_3$ with total energy $M$ in the center-of-mass system. The exact expression is rather complicated (see Eq. (3.18) in Ref. \cite{Meggiolaro2011}, and also Ref. \cite{DD2004} and references therein), but it is surely a good approximation to take $m_a\simeq0$, considering the experimental upper bound on the axion mass $m_a\lesssim 10^{-2}$ eV \cite{bounds,IR2018}. The expression for the phase space for two particles of mass \textit{m} and one massless particle turns out to be
\begin{equation} 
\begin{split}
&\Phi^{(3)}(M|m,m,0) = \\
&\frac{M^{2}}{256\pi^{3}}\left\{\left(1+\frac{2m^{2}}{M^{2}}\right)\sqrt{1-\frac{4m^{2}}{M^{2}}}-\frac{4m^{2}}{M^{2}}\left(1-\frac{m^{2}}{M^{2}}\right)\ln\left[\frac{M^{2}}{2m^{2}}\left(1+\sqrt{1-\frac{4m^{2}}{M^{2}}}\right)-1\right]\right\} .
\end{split}
\end{equation}
Inserting the numerical values of the constants \eqref{masse}--\eqref{masses} and $F_\pi$, we obtain the following results:
\begin{equation} \label{decayw}
\begin{cases}
\Gamma(\eta\rightarrow\pi^{0}\pi^{0}a)=b^{2}(5.62\pm0.04)\times10^{-3} ~\text{MeV} ,\\
\Gamma(\eta\rightarrow\pi^{+}\pi^{-}a)=b^{2}(10.52\pm0.07)\times10^{-3} ~\text{MeV} ,\\
\Gamma(\eta'\rightarrow\pi^{0}\pi^{0}a)=b^{2}(2.49\pm0.02)\times10^{-2} ~\text{MeV} ,\\
\Gamma(\eta'\rightarrow\pi^{+}\pi^{-}a)=b^{2}(5.01\pm0.03)\times10^{-2} ~\text{MeV} .
\end{cases}
\end{equation}

\newsection{Conclusions: summary of the results and prospects}

In this paper we have investigated the most interesting decay processes involving axions, photons and the lightest pseudoscalar mesons, making use of the ``axionized'' version of a chiral effective Lagrangian model proposed by Witten, Di Vecchia, Veneziano, \textit{et al.} (WDV), which describes the Nambu-Goldstone bosons originated by the spontaneous breaking of the $SU(3)_L\otimes SU(3)_R$ chiral symmetry (with $L=3$ light quark flavors) and the flavor-singlet pseudoscalar meson, implementing the $U(1)$ axial anomaly of the fundamental theory.

In particular, in Sec. 3 we have computed the axion-photon-photon coupling constant $g^{QCD}_{a\gamma\gamma}$ and the result is given by the expression \eqref{axioncoupling}, that we have compared with the one obtained using the Chiral Effective Lagrangian with $L=2$ light quark flavors.
As we have verified in Sec. 5, this expression (as well as the expression for the mass of the axion) is not modified using, in place of the WDV model, a ``natural'' extension of it which also includes an extra U(1) axial condensate.

In Table 1 of Sec. 6 we have reported the numerical estimate for the axion-photon-photon coupling constant \eqref{axioncoupling}: for comparison, we have also reported the corresponding estimates derived using the Chiral Effective Lagrangian with $L=2$ light quark flavors at LO and NLO.
Comparing our result with the estimate found using the $L=2$ Chiral Effective Lagrangian at LO, we get a value which is about 9\% smaller, and it is also a bit smaller than (but almost compatible within the errors with) the value obtained using the $L=2$ Chiral Effective Lagrangian at NLO.
Of course, in the hypothesis that this type of process will be observed in the future, it will be important to know the level of accuracy of a given theoretical estimate, when comparing it with the experimental result, and in this perspective our result will be surely relevant.
Looking at the values reported in Table 1, one could optimistically consider our result as a more ``precise'' determination, with respect to the result obtained using the $L=2$ Chiral Effective Lagrangian at LO and NLO (because, maybe, our effective model, already at tree level, is able to reproduce results with an accuracy comparable to the one which is obtained using the $L=2$ Chiral Effective Lagrangian after including many higher-order corrections\dots).
Adopting, instead, a more conservative approach, one could simply consider our result as an alternative determination using a chiral effective Lagrangian model, which (when compared with other similar determinations) allows to estimate a sort of ``systematic uncertainty'' for this kind of theoretical predictions.

Then, in Sec. 4 we have performed an explorative study of the hadronic decays $\eta/\eta'\rightarrow\pi\pi a$ (which, among all the possibile hadronic decays involving also the axion, are the ones involving the lowest-energy hadrons):
the expressions for the amplitudes $g_{\eta\pi\pi a}$ and $g_{\eta'\pi\pi a}$ are given by Eqs. \eqref{couplingh1} and \eqref{couplingh2} respectively.
In Sec. 6, Eq. \eqref{decayw}, we have reported the numerical estimates for the corresponding decay widths: these are the main original results obtained in this paper. (Moreover, as we have found in Sec. 5, the addition of a possible U(1) axial condensate, while not modifying the $\eta\rightarrow\pi\pi a$ decays, makes the $\eta'\rightarrow\pi\pi a$ decay widths a bit smaller by a factor $ 0.88 \lesssim (F_\pi/F_{\eta'})^2 \leq 1$.)
Considering the existing experimental bounds on $b$ (based on astrophysical and cosmological considerations) \cite{bounds}, $ 10^{-18} \lesssim b\lesssim 10^{-10}$, the decay widths \eqref{decayw} turn out to very small, smaller than about $10^{-22}$ MeV.\footnote{Just for comparison, we recall here the experimental bounds on the CP-violating $\eta/\eta'$ decays in two pions \cite{PDG}:
$\Gamma^{exp}(\eta\rightarrow\pi^{0}\pi^{0})<4.6\times10^{-7}$ MeV,
$\Gamma^{exp}(\eta\rightarrow\pi^{+}\pi^{-})<1.7\times10^{-8}$ MeV,
$\Gamma^{exp}(\eta'\rightarrow\pi^{0}\pi^{0})<7.8\times10^{-5}$ MeV,
$\Gamma^{exp}(\eta'\rightarrow\pi^{+}\pi^{-})<3.5\times10^{-6}$ MeV.
We also observe that, even with the largest value of $b$ allowed by the above-mentioned astrophysical bounds, i.e, $b \simeq 10^{-10}$, the decay widhts \eqref{decayw} turn out to be about a factor $10^{-3}$ (for $\eta\rightarrow\pi\pi a$) and $10^{-2}$ (for $\eta'\rightarrow\pi\pi a$) smaller than the model-independent bounds on the rates of the rare (CP-violating) decays $\eta(\eta')\rightarrow\pi\pi$, which have been derived in Ref. \cite{cpviolating}, using the experimental limits on the neutron electric dipole moment.}
As far as we know, no experimental search for these processes has been attempted up to know. However, even if the electromagnetic decay of the axion ($a\rightarrow\gamma\gamma$) certainly remains the most promising process which might provide some experimental signature of the axion, we believe that it would be worthwhile to look also for these possible decay processes $\eta/\eta'\rightarrow\pi\pi a$ in future $\eta$-factory experiments.

We conclude by observing that our estimates \eqref{decayw} for the widths of the $\eta/\eta'\rightarrow\pi\pi a$ decays are based on the expressions \eqref{couplingh1} and \eqref{couplingh2} for the amplitudes $g_{\eta\pi\pi a}$ and $g_{\eta'\pi\pi a}$, which have been obtained \emph{directly} (i.e., at \emph{leading order}) from our chiral effective Lagrangian model (described in Sec. 2).
It is plausible that these LO estimates will receive large contributions from chiral loop corrections at NLO and NNLO, and from strong final-state rescattering (as it happens, for example, in the $\eta/\eta'\rightarrow 3\pi$ decays).
Alternatively, one could consider the approach described in Refs. \cite{ASW2019,FS1999}, in which one takes into account an \emph{extended} chiral effective Lagrangian model, which also includes the lowest-lying nonet of scalar mesons (i.e., a \emph{linearized} version of the [nonlinear] chiral effective Lagrangian model described in Sec. 2): in this alternative approach, also contributions to the $\eta/\eta'\rightarrow\pi\pi a$ amplitudes coming from scalar-meson exchanges are taken into account.\\
We believe that it would be worthwhile to go beyond the explorative study undertaken in this paper and to better investigate the $\eta/\eta'\rightarrow\pi\pi a$ decays following the ``guidelines'' mentioned above: some progress in these directions is expected in the near future.

\newpage

\renewcommand{\Large}{\large}

\end{document}